\documentclass[aps,prl,twocolumn,superscriptaddress,nofootinbib,noshowpacs,10pt]{revtex4-1}
\usepackage[]{amsmath}
\usepackage{graphics}
\usepackage{amssymb}
\usepackage{graphicx}
\usepackage{color}
\usepackage{bm}

\newcommand{\be}{\begin{equation}}
\newcommand{\ee}[1]{\label{#1} \end{equation}}

\newcommand {\ket}[1]{\lvert \, #1\rangle}
\newcommand {\bra}[1]{\langle #1 \, \rvert}
\newcommand {\braket}[2]{\langle #1 \, | \, #2 \rangle}

\def\a{\hat{a}}
\def\A{\hat{a}^{\dagger}}

\def\f12{\frac{1}{2}}

\newcommand{\mean}[1]{\ensuremath{\left\langle #1 \right\rangle}}

\newcommand {\comm}[2]{\left[ #1 , #2 \right]}

\newcommand {\hp}{\ensuremath{\hat{\Phi}}}

\newcommand {\dhp}{\ensuremath{\dot{\hat{\Phi}}}}

\newcommand {\ddhp}{\ensuremath{\ddot{\hat{\Phi}}}}

\begin{document}
\title{Quantum coherent oscillations in the early universe}
\date{\today}

\author
{Igor Pikovski}
\email{igor.pikovski@cfa.harvard.edu}
\affiliation{ITAMP, Harvard-Smithsonian Center for Astrophysics, Cambridge, MA 02138, USA}
\affiliation{Department of Physics, Harvard University, Cambridge, MA 02138, USA}
\author
{Abraham Loeb}
\email{aloeb@cfa.harvard.edu}
\affiliation{ITC, Harvard-Smithsonian Center for Astrophysics, Cambridge, MA 02138, USA}
\affiliation{Department of Astronomy, Harvard University, Cambridge, MA 02138, USA}


\begin{abstract}
Cosmic inflation is commonly assumed to be driven by quantum fields. Quantum mechanics predicts phenomena such as quantum fluctuations and tunneling of the field. Here we show an example of a quantum interference effect which goes beyond the semi-classical treatment and which may be of relevance in the early universe. We study the quantum coherent dynamics for a tilted, periodic potential, which results in genuine quantum oscillations of the inflaton field, analogous to Bloch oscillations in condensed matter and atomic systems. Our results show that quantum interference phenomena may be of relevance in cosmology.

\end{abstract}

\maketitle

It is commonly assumed that the universe underwent inflation, an early epoch of rapid expansion \cite{1980PhLB...91...99S, 1981PhRvD..23..347G,  1982PhLB..108..389L, 1982PhRvL..48.1220A}. Inflation successfully explains a number of cosmological observations, such as the near homogeneity and isotropy of the universe. In addition, quantum fluctuations during inflation are assumed to have seeded the observed large-scale inhomogeneities in the matter distribution. There is currently much interest in identifying other observable signatures of genuine quantum effects in cosmology \cite{2014PhLB..739..285C, 2015arXiv150206770H, 2015arXiv150308043A, 2015arXiv150801082M}. In this {\it Letter}, we treat the inflaton field quantum mechanically and study its quantum dynamics. Due to quantum interference phenomena, the dynamics can significantly differ from the semi-classical predictions, with possible consequences for observations.

In the simplest model of inflation a single scalar field is responsible for the rapid expansion of the universe. Here we briefly review the main results \cite{mukhanov2007introduction, dodelson2003modern}. The classical action for the scalar field $\Phi$ with a potential $V(\Phi)$ and gravity is
\be
S = \!  \int \!  d^4x \sqrt{|g|} \left( \frac{m_P^2 }{2}R - \f12 g^{\mu \nu} \nabla_{\mu}\Phi \nabla_{\nu}\Phi - V(\Phi) \right) ,
\ee{eq:action}
where $R$ is the Ricci-scalar, $g$ the determinant of the metric and $m_P = (8 \pi G)^{-1/2}$ is the reduced Planck-mass (we use units with $\hbar = c =1$ throughout the manuscript).
The homogeneous and isotropic solution for the gravitational field is the FRW-metric, given by
\be
ds^2 = - dt^2 + a(t)^2 \left( dr^2 + r^2 d\theta^2 + r^2 \textrm{sin}^2\theta d\phi^2 \right) ,
\ee{eq:FRW}
where for simplicity a spatially flat universe was assumed. Defining $H= \dot{a}/a$, where the dot denotes differentiation with respect to the time coordinate $t$, the Einstein equations for a homogeneous field $\Phi(\mathbf{x},t)= \Phi(t)$ yield
\be
\begin{split}
H^2 & = \frac{1}{3 m_P^2} \left( \f12 \dot{\Phi}^2 + V(\Phi)\right) \\
\dot{H} & = -\frac{1}{2 m_P^2} \dot{\Phi}^2
\end{split}
\ee{eq:Hubble}
The Klein-Gordon equation on the above metric becomes
\be
\ddot{\phi} + 3 H \dot{\Phi} + V'(\Phi) =0 ,
\ee{eq:Phi}
where the prime denotes differentiation with respect to $\Phi$.
The Hubble parameter $H$, characterizing the expansion of the universe, depends on the field $\Phi$ and its potential $V(\Phi)$.
These cause the scale factor $a(t)$ to grow exponentially during most of the inflationary period.
Towards the end of inflation the potential energy is reduced, the ``friction'' term $3 H \dot{\Phi}$ in eq. \eqref{eq:Phi} can become negligible and the field oscillates inside the potential $V(\Phi)$. As the expansion ends, the field decays into other particles. This can be approximately captured by an additional effective damping term $\Gamma \dot{\Phi}$ in the equation of motion \eqref{eq:Phi}.

The equations of motion are analogous to the description of a particle in a potential. Despite the use of a quantum field in the model, the description so far was completely classical. In the usual treatment, inhomogeneous perturbations to the field, $\delta \Phi(\mathbf{x},t) $, are considered in the linearized regime and quantized. These perturbations, which are interpreted to arise due to vacuum fluctuations, become squeezed due to inflation and decay into the matter distribution we see today.

Here we consider the quantum dynamics of the \textit{full} inflaton field. To this end, we consider the equation of motion \eqref{eq:Phi} for the quantized homogeneous field $\hat{\Phi}$ and study the resulting quantum dynamics. We consider standard quantization with $\comm{\hat{\Phi}(\mathbf{x})}{\hat{\pi}(\mathbf{y})}= i \hbar \delta(\mathbf{x}- \mathbf{y})$, where $\hat{\pi}$ is the conjugate momentum field operator.
In the usual, semi-classical treatment, the field has always a specific well-defined value, say $\Phi_1$. In this treatment, quantum tunnelling can transform the field from one well-defined value to another, $\Phi_1 \rightarrow \Phi_2$ \cite{1977PhRvD..15.2929C}. The system can probabilistically ``penetrate'' a classical barrier, but it materializes with a classically well defined value afterwards. However, quantum mechanically the field can be in a superposition of different amplitudes, $\ket{\Phi_1}+\ket{\Phi_2}$, which is not captured by a semi-classical treatment. Until a measurement is performed, or decoherence occurs, superpositions of different amplitudes are physically valid solutions. Usually, such states are fragile and would quickly decohere, as for example in condensed matter systems. Within the single-field inflationary model, however, one can expect quantum coherence to be preserved. Assuming that no deleterious effects destroy the quantum coherence until some time $t^*$, the field can evolve into quantum superpositions for times $t<t^*$. For a quantum state, one can only assign probabilities to each specific value of the field, but none of the possible superposed outcomes are yet classically realized.

In our discussion the quantized nature of gravity is neglected. In other words the gravitational degrees of freedom remain fully classical and we interpret $H=\mean{H}$ as an effectively classical variable.
Clearly such a treatment is incomplete in describing the full evolution during all of inflation and the entanglement with gravitational degrees of freedom is not accounted for. To incorporate the latter, a quantum theory of gravity would be necessary, which is outside the scope of this work. As the field evolves during inflation, the ``damping'' term $3H\dot{\Phi}$ also suppresses the fragile quantum dynamics. Nonetheless, one can expect our treatment to be approximately valid when the dynamics of the metric is slow, i.e. either before or after slow-roll inflation.

We are mostly interested in the time evolution of the quantum scalar field when the damping due to expansion can be neglected. In this regime the quantum dynamics of the field is governed by
\be
\ddhp + V'(\hat{\Phi}) =0 \, .
\ee{eq:PhiQ}
Rather than studying the decomposition of the field into individual modes, we consider the equation of motion for the entire field. This resembles the quantum evolution of a system in first quantization. Note that the dynamics of quantum systems in the presence of potentials has been extensively studied in low-energy quantum theory \cite{2009arXiv0906.1640P}. Here we can directly apply some results to the field operator $\hat{\Phi}$, since the equation of motion is analogous to the Heisenberg equation of motion for the position operator in non-relativistic quantum theory.

\begin{figure}[t]
\centering
\includegraphics[width=0.9\columnwidth]{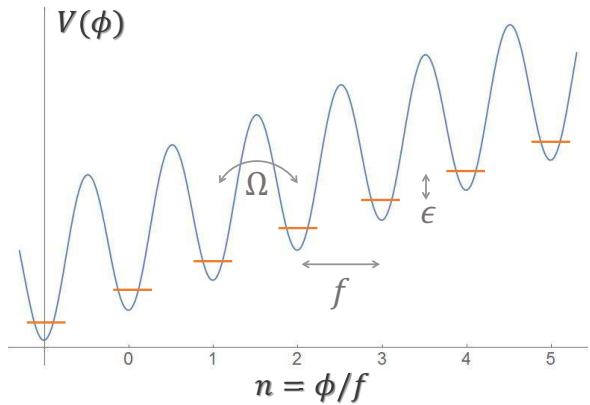}
\caption{\small The assumed potential for $\Phi$, which consists of a periodic part and an additional homogeneous field (see eq. \eqref{eq:Abbott}). The minima (sites) are labeled by $n$, the orange lines represent the local ground states. The field can tunnel from one site to another, governed by the rate density $\Omega$. Due to the tilt in the potential, the quantum dynamics results in coherent, periodic oscillations over many sites.}
\label{fig_plot1}
\end{figure}

In particular, we consider the model by Abbott \cite{1985PhLB..150..427A} which considers a potential of the form
\be
V (\hp) = V_0 \cos\left( \frac{2 \pi \hp}{f}\right) + \epsilon \frac{\hp}{f} ,
\ee{eq:Abbott}
as also depicted in Figure \ref{fig_plot1}. We assume this potential describes the dynamics in the regime that we are interested in, namely towards the beginning and end of inflation. This model has also been studied in the context of a cyclic universe \cite{2002Sci...296.1436S}.
The quantum dynamics for this type of potential is very different from the classical counterpart: quantum systems undergo coherent, periodic oscillations around the classical mean.  The eigenstates for this potential are not localized inside a potential well; rather they are delocalized over several minima. Thus, if the system is initially inside a well, it will not have a well-defined energy but will be in a superposition of different energy eigenstates. The system will evolve in time and spread out (this being also the physical reason for quantum tunneling). In other words, the states localized in the minima are not ground states or true vacuum states, but will slowly delocalize. Importantly, the delocalization is also altered by the linear contribution to the potential: the ``tilt'' is responsible for coherent oscillations of the system, also known as Bloch oscillations \cite{1929ZPhy...52..555B, 1934RSPSA.145..523Z}. These oscillations can extend over many minima and are quantum mechanical in nature, as they arise due to quantum interference. Such oscillations have been observed in various experiments \cite{1993PhRvL..70.3319W, 1996PhRvL..76.4508B, 2015Sci...347.1229P}.

To study the effect in detail, we can directly use the developed techniques in low-energy quantum theory \cite{1973PhRvB...8.5579F, 1996PMagB..74..105H, 2004NJPh....6....2H, 2003PhLA..317...54K}. It is convenient to express the full dynamics of the system with potential \eqref{eq:Abbott} in the Hamiltonian formulation, where the dynamics of $\hp$, eq. \eqref{eq:PhiQ}, is generated by an effective Hamiltonian $\hat{H} = \int d^3x \hat{\cal{H}}$ with the Hamiltonian density
\be
\hat{\cal{H}} = -  \frac{\Omega}{4} \sum_{n= - \infty}^{\infty} (\a_n \A_{n+1} + \a_{n+1} \A_n ) + \epsilon \sum_{n= - \infty}^{\infty} n \A_n \a_n .
\ee{eq:H}
Here, the operator $\A_n $ ($\a_n$) creates (annihilates) a field excitation in the $n$-th local potential well minimum (see Figure \ref{fig_plot1}), and $\Omega $ is the hopping rate per unit volume between neighboring sites. In terms of these operators, the field is $\hp =f \sum_n n \A_n \a_n$. The use of the above Hamiltonian assumes only nearest-neighbor hopping, and neglects any higher energy levels inside the minima (only the local false vacuum states are considered).

In the Heisenberg picture, the operator $\hp(t)$ evolves in time according to eq. \eqref{eq:PhiQ}. The same Hamiltonian can also be used to study the dynamics in the Schroedinger picture \cite{1992PhRvD..45.2044K, 2013PhRvD..88h5020D}, i.e. the dynamics of the state, or wave functional, is given by $\ket{\Psi(t)} = \hat{U} \ket{\Psi(0)} =e^{-i \hat{H} t } \ket{\Psi(0)}$. The Schroedinger picture can be instructive to gain insight into the quantum processes that arise.

\begin{figure}[t]
\centering
\includegraphics[width=0.85\columnwidth]{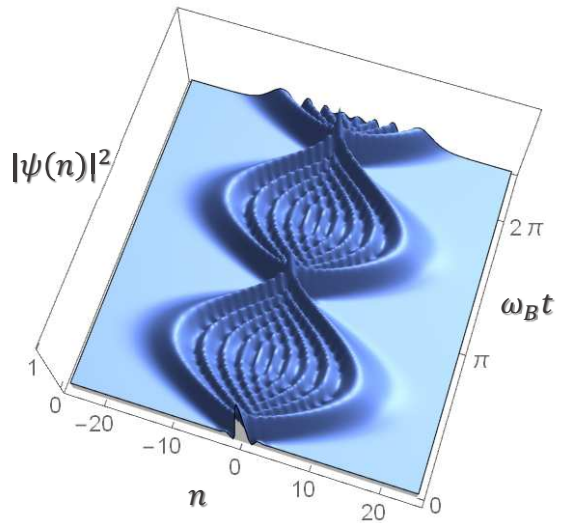}
\caption{\small Bloch oscillations of the system, for an initially localized state inside a potential well. The state periodically spreads over many sites of the potential (labeled by $n$), with frequency $\omega_B$. The maximum amplitude of the oscillations is $l \approx \omega / \omega_B$.}
\label{fig_Bloch1}
\end{figure}

Classically, the system would be accelerated by the linear term in eq. \eqref{eq:Abbott}, but if trapped in a potential minimum, it would remain bound. Semi-classically, the system can tunnel to the neighboring site, governed by the rate $\omega = \int d^3x \Omega$. The full quantum dynamics given by eq. \eqref{eq:H} predicts a coherent quantum phenomenon: Bloch oscillations of the system at a frequency $\omega_B = \int d^3 x \epsilon$. These oscillations can be described in terms of the eigenstates of the Hamiltonian \eqref{eq:H}, the  Wannier-Stark states \cite{1960PhRv..117..432W} $\ket{\psi_m}$ with eigenenergies $E_m = m \epsilon$ with integer $m$. In terms of the localized states $\ket{\Phi_n} = \ket{n} $ corresponding to site $n$, these are given by
\be
\ket{\psi_m} = \sum_n J_{n-m}\left(\frac{\omega }{2 \omega_B}\right) \ket{n} ,
\ee{eq:WS}
where $J_k$ are Bessel functions of the first kind.
Using the summation property of the Bessel function, $\sum_n J_n(x) J_{n+k}(x) e^{2iky}=i^k J_k(2x \,  \textrm{sin}y)e^{-ikx/2}$, one finds the propagator from site $k$ to $n$ to be
\be
\begin{split}
& \bra{n} \hat{U} \ket{k}  = \sum_j  e^{-i E_j t} \braket{n}{\psi_j} \braket{\psi_j}{k} \\
& = J_{n-k}\left(\frac{\omega}{\omega_B} \sin(\frac{ \omega_B t}{2})\right) e^{i \f12(n+k)\omega_B t} i^{n-k} .
\end{split}
\ee{eq:propagator}
Thus, if the system is initially localized at some site (or in other words, is in a false vacuum)$\ket{\Psi(0)}=\ket{n=0}$, the time evolution is
\be
\ket{\Psi(t)} = \sum_n i^n J_n\left(\frac{\omega}{\omega_B} \sin(\frac{ \omega_B t}{2})\right) e^{-\frac{i}{2}\omega_B t} \ket{n} .
\ee{eq:time}
%
During the full evolution, the mean remains unaffected  ($\mean{\Phi} =0$), but the system coherently spreads periodically over many sites, with the Bloch frequency $\omega_B$. The largest oscillation amplitude of the field is on the order of $L \approx f \omega / \omega_B $. After a time $T_B = 2 \pi / \omega_B$, the state is given by its initial value, $\ket{\Psi(T_B)}=\ket{\Psi(0)}$, see Figure \ref{fig_Bloch1}. At each instant of time, one can only assign a probability to a certain field value, given by $|\braket{n}{\Psi(t)}|$. The state remains delocalized until a measurement or decoherence occurs.

One can also find the solutions in the Heisenberg picture \cite{1996PMagB..74..105H, 2003PhLA..317...54K}, i.e. the time evolved field state $\hp(t)$. To this end, we follow the algebraic approach as developed in Ref. \cite{2003PhLA..317...54K}. The displacement operator $\hat{D}$ shifts the field from one site to the next, i.e. $\hat{D} \ket{n} = \ket{n+1}$ and $\hat{D}^\dagger \ket{n} = \ket{n-1}$, and has the properties $\hat{D} \hat{D}^{\dagger} = 1$, $\comm{\hat{D}}{\hp}=f \hp$. The Heisenberg equation of motion for the displacement operator gives the solution $\hat{D}(t) = \hat{D}(0) e^{i \omega_B t}$. In terms of the displacement operator, the field operator evolves in time as
\be
\dhp(t) = \frac{i f \omega }{4} \left( \hat{D}(0) e^{i \omega_B t} - \hat{D}^\dagger(0) e^{-i \omega_B t} \right)
\ee{eq:phidt}
with the solution
\be
\hp(t) = \hp(0) + \frac{ f \omega }{4 \omega_B} \left( \hat{D}(0) \left( e^{i \omega_B t} -1 \right) + \hat{D}^\dagger(0) \left( e^{-i \omega_B t} -1 \right) \right)
\ee{eq:phit}

We link the results to cosmological parameters via eq. \eqref{eq:Hubble}, interpreting $H$ as a semi-classical mean value.
If the field has initial Gaussian spread over many sites, $\ket{\Psi(0)}=\mathcal{N} \sum_n e^{-n^2/2\sigma^2 }\ket{n}$, where $\mathcal{N}$ is the normalization factor, the relevant quantity becomes
\be
\mean{\dhp^2} = \frac{f^2 \omega^2}{8} \left(1- e^{-1/\sigma^2} \textrm{cos}( 2 \omega_B t) \right) \approx \frac{f^2 \omega^2}{4} \textrm{sin}^2(\omega_B t)  ,
\ee{eq:2}
where the last approximation holds for an initially broad Gaussian. For this case, the quantity $\mean{\dhp^2}$ undergoes oscillations at frequency  $\omega_B$.
We thus have
\be
\begin{split}
H^2 =  \frac{f^2 \omega^2}{24 m_P^2} \textrm{sin}^2(\omega_B t) + \frac{1}{3 m_P^2} \mean{V(\Phi)} 
\end{split}
\ee{eq:Hubble2}
In the case when $\mean{\dhp^2}$ is comparable to $ \mean{V(\Phi)}$ (so for the occupied sites $n$ sufficiently low in the potential landscape such that the potential energy is comparable to the kinetic energy, $n \sim f^2 \omega^2/\epsilon$), the quantum coherent oscillations of the inflaton field can significantly contribute to the value of the Hubble parameter.

We note that in deriving the results, the term $ 3 H \dot{\Phi}$ in eq. \eqref{eq:Phi} has been neglected. One can effectively include the term by inserting the result from eq. \eqref{eq:Hubble2} back in eq. \eqref{eq:Phi}. This term will cause the Bloch oscillations to be damped on a time-scale of order $1/H$.
Neglecting the damping term can hold during ``fast roll'', when the acceleration dominates over the friction term (for the example considered here, this constitutes the regime $H \ll \omega_B$ or $1 \gg 3H\dot{\Phi}/\ddot{\Phi} \approx f \omega/(m_{P}\omega_B)$,  requiring that the periodicity of the field is smaller than the Planck-mass). One can expect this condition to hold at the early stages of inflation and towards the end of inflation, thus one may expect the strongest Bloch oscillations to occur then. This translates to super-horizon scales and small scales, respectively. In addition to the damping, any possible decoherence channel will destroy quantum coherence and the associated Bloch oscillations \cite{2003PhRvL..91y3002B}. Within the single-field inflationary model, however, one would expect quantum coherence to be preserved.

Linking our results to observations of the cosmic microwave background anisotropies or large scale structure requires the study of spatial perturbations on top of the quantum solution for the homogeneous field, which will be studied elsewhere. Oscillations such as in eq. \eqref{eq:Hubble2} would correspond to periodic changes in the timing at which field fluctuations exit the horizon. This would lead to a periodic modulation of the power spectrum of matter density fluctuations in the observable universe. Their detection could provide evidence of quantum coherent phenomena during inflation (although a different potential could lead to classical oscillations with a similar signature; discerning the quantum oscillations from classical ones requires knowledge of the potential).

The quantum coherent oscillations discussed in this {\it Letter} have been derived for the specific, periodic Abbott potential \cite{1985PhLB..150..427A}. The present analysis serves as an example for quantum interference effects that may be of relevance in cosmology. The example of Bloch oscillations may in fact be relevant whenever several potential minima are present. For Bloch oscillations to occur, the boundary conditions are negligible as long as they are further away than the maximum coherent oscillation amplitude $L=f\omega/\omega_B$ \cite{1973PhRvB...8.5579F}. Thus, for a potential with many local minima and periodicity over a finite region, similar behaviour can be expected. In string theory for example, one expects a vast potential landscape \cite{2003JHEP...05..046D} where such conditions may be met. If, however, the minima are completely disordered and spread out randomly, quantum interference results in localization of the field in analogy to Anderson localization \cite{1958PhRv..109.1492A}.

Finally, we note that coherent and periodic oscillations of the field may be of interest in the context of ``doomsday'' scenarios \cite{2015arXiv150308130S, 1980PhRvD..21.3305C}, in which bubble nucleation at different local potential minima may occur.

\vspace{3pt}
\textit{Acknowledgements.}
We thank Cliff Burgess, Xingang Chen, James Hartle, Juan Maldacena, Richard Schmidt and Alexander Vilenkin for discussions.
This work was supported in part by NSF through grant AST-1312034 and through a grant to ITAMP.

\bibliography{cosbib}

\end{document}